\documentclass[11pt]{iopart}

\usepackage{iopams}  

\usepackage{color}
\usepackage{hyperref}
\usepackage[usenames,dvipsnames,svgnames,table]{xcolor}
\usepackage{ulem}

\usepackage{mathrsfs} % script math
\usepackage{ulem} % for \sout \xout....

\begin{document}

\paper{Boosted Schwarzschild Metrics from a Kerr-Schild Perspective }

\author{Thomas M\"adler$^{1,2}$\footnote{Email:thomas.maedler@mail.udp.cl} and Jeffrey Winicour$^{3,4}$}

\address{
${}^1$ N\'ucleo de Astronom\'ia, Facultad de Ingenier\'ia, Universidad Diego Portales, Av. Ej\'ercito 441, Santiago, Chile\\
${}^{2}$  Institute of Astronomy, University of Cambridge, Madingley Road, Cambridge, CB3 0HA, UK \\
${}^{3}$ Department of Physics and Astronomy \\
        University of Pittsburgh, Pittsburgh, PA 15260, USA\\
${}^{4}$ Max-Planck-Institut f\" ur
         Gravitationsphysik, Albert-Einstein-Institut, \\
	 14476 Golm, Germany \\
	}

\begin{abstract}

The Kerr-Schild version of the Schwarzschild metric contains
a Minkowski background which provides a definition
of a boosted black hole. There are two Kerr-Schild versions
corresponding to ingoing or outgoing principle null directions.
We show that the two corresponding Minkowski backgrounds
and their associated boosts
have an unexpected difference. We analyze this difference and
discuss the implications in the nonlinear regime
for the gravitational memory effect resulting from the
ejection of massive particles from an isolated system. 
We show that the nonlinear effect agrees with the linearized result based upon the retarded Green function only if the velocity of the ejected particle corresponds
to a boost symmetry of the ingoing Minkowski
background. A boost with respect to the outgoing
Minkowski background is inconsistent with the absence
of ingoing radiation from past null infinity.

\end{abstract}


%\maketitle
%\tableofcontents

\section{Introduction}

By considering retarded solutions of the linearized Einstein
equation on a Minkowski background, Zeldovich and
Polnarev~\cite{zeld}
pointed out the existence of a memory effect in the gravitational
waves produced by the ejection of massive particles
to infinity.
Our previous work~\cite{boost} has shown that this
effect could also be obtained in linearized theory
by considering the transition from
an initial state whose exterior was described by
a Schwarzschild metric at rest to a final state
whose exterior was a boosted Schwarzschild metric.
The results
were based upon a Kerr-Schild version of the
Schwarzschild metric to describe the far field exterior
to what we referred to as a Schwarzschild body.
For such a body
in linearized theory which is initially at rest,
then goes through a radiative stage and finally emerges
in a boosted state, we showed that
the proper retarded solution for the resulting memory
effect is described in terms of the ingoing version of
the Kerr-Schild metric for both the initial and final states.
An outgoing Kerr-Schild or time symmetric Schwarzschild metric
would give the wrong result.
The result was independent of the details of the intervening
radiative period. Because the Kerr-Schild metrics
are solutions both in the linearized and nonlinear
sense, we extrapolated this result to the nonlinear
case.

Here, we investigate this problem from
the purely nonlinear perspective. There are
two major differences from the linearized view.

The first major difference is that the linearized result
in~\cite{boost} was obtained using the boost
associated with the Lorentz symmetry of the unperturbed
Minkowski background. The Kerr-Schild metrics~\cite{ks1,ks2}
have the form
\begin{equation}
    g_{\mu\nu} = \eta_{\mu\nu} +H \ell_\mu \ell_\nu
    \label{eq:ksk}
\end{equation}
where $\eta_{\mu\nu}$ is a Minkowski metric,
$\ell_\mu$ is a principle null vector field (with respect to
both $\eta_{\mu\nu}$ and $g_{\mu\nu}$)
and $H$ is a scalar function. In the nonlinear case,
there are two natural choices of ``Minkowski background''
$\eta_{\mu\nu}$ depending on whether the null vector
$\ell^\mu$ in the Kerr-Schild metric (\ref{eq:ksk})
is chosen to be in the ingoing or
outgoing direction.
\footnote{Note that a time symmetric version of
the Schwarzschild metric, see \eref{eq:Schwarz_std}, does not single
out such a preferred Minkowski background
in the nonlinear case.}.  

The second major difference in the nonlinear case
is that there is no analogue of the Green function to construct
a retarded solution. Instead, the retarded solution due
to the emission of radiation from an accelerated particle
is characterized by the absence of ingoing radiation
from ${\mathcal I}^-$. A necessary condition that there be no ingoing
radiation is that the analogue of the radiation memory at past
null infinity ${\mathcal I}^-$ vanishes.
In that case the ingoing radiation
strain, which forms the free characteristic initial data on
${\mathcal I}^-$, may be set to zero.
Otherwise, as explained in Sec.~\ref{sec:mem},
non-vanishing radiation memory at  ${\mathcal I}^-$
requires that there must ingoing radiation from ${\mathcal I}^-$.

Consequently, since an initial unboosted Kerr-Schild-Schwarzschild
metric has vanishing radiation strain 
at ${\mathcal I}^-$, the final boosted metric must also
have vanishing radiation strain at ${\mathcal I}^-$ if
there is no intervening ingoing radiation.
This is the case if the boost belongs to the Lorentz subgroup of the
BMS group at ${\mathcal I}^-$.
This corresponds to the boost
symmetry of the Minkowski metric associated with the
ingoing version of the Kerr-Schild metric. 
On the contrary, a boost with respect to the Minkowski
metric associated with the outgoing version of the
Kerr-Schild metric leads to 
non-vanishing radiation memory at ${\mathcal I}^-$
so that it is inconsistent with the requirement of
vanishing ingoing radiation.

This leads to our main result: the calculation in the nonlinear regime
of the memory effect due to the ejection of
a massive particle is correctly described by the boost
associated with Minkowski background of the ingoing
Kerr-Schild metric. The key ingredient is that this
boost is a BMS symmetry at ${\mathcal I}^-$
but not at ${\mathcal I}^+$. This leads
to vanishing radiation memory at ${\mathcal I}^-$
but to non-zero radiation memory at ${\mathcal I}^+$, which
is in precise agreement
with the extrapolation from the linearized
result based upon the retarded Green function.

In Sec.~\ref{sec:advret}, we discuss unexpected
features which result in defining the boost in terms of
the Lorentz
symmetries of the Minkowski backgrounds of either the
ingoing or outgoing versions of the Kerr-Schild metric.
This requires considerable notational
care, which  warrants the formalism presented in
Sec.~\ref{sec:advret}. We show that the boost symmetry
of the Minkowski metric associated with the ingoing
version of the Kerr-Schild metric corresponds
to the boost symmetry of the Bondi-Metzner-Sachs
(BMS)~\cite{Sachs_BMS} asymptotic symmetry group at
${\mathcal I}^-$ but is a singular transformation
at future null infinity ${\mathcal I}^+$. Conversely, the boost symmetry
of the Minkowski metric associated with the outgoing version
of the Kerr-Schild metric corresponds to the boost symmetry of the
BMS group at ${\mathcal I}^+$ but is a singular transformation
at ${\mathcal I}^-$.

The Kerr-Schild metrics have played an important
role in the construction of exact solutions; see~\cite{exact}.
The most important examples are the Schwarzschild
and Kerr black hole metrics. Because their metric form
(\ref{eq:ksk}) is
invariant under the Lorentz symmetry of the
Minkowski background $\eta_{\mu\nu}$, the boosted Kerr-Schild versions of the
Schwarzschild and Kerr metrics have also played an important role in
numerical relativity in prescribing initial data for superimposed black
holes in a binary orbit~\cite{ksm1,ksm2}. The initial
data for the numerical simulations are prescribed
in terms of the ingoing version of
the Kerr-Schild metric, whose coordinatization in terms
of advanced time covers the interior of
the future event horizon. The initial velocities of the
black holes are generated by the boost symmetry of 
the Minkowski background for the ingoing version of the
Kerr-Schild metric. Surprisingly, this
boost symmetry, which is a well-behaved BMS transformation
at ${\mathcal I}^-$, has singular behavior
at ${\mathcal I}^+$. This overlooked asymptotic
property of the boost symmetry could possibly 
introduce spurious asymptotic
behavior in the Kerr-Schild construction of binary black
hole initial data.

We concentrate here on
the boosted Schwarzschild metric. The Kerr case is more
complicated because the twist of the principle
null direction $\ell_\mu$ does not allow a
straightforward construction of
well-behaved advanced or retarded null coordinate
systems. Although an exact Schwarzschild exterior is
unrealistic in a dynamic spacetime it is reasonable to
expect that our
results are valid if it is a good far field approximation
in the neighborhood of null infinity
in the limit of both infinite future and infinite
past retarded and advanced times. In this respect, our results might
also apply to the Kerr case since the metric terms involving
the angular momentum parameter fall off faster
with $r$ than the terms involving the mass.

\section{Boosts and the Kerr-Schild-Schwarzschild metrics}
\label{sec:advret}

In time symmetric coordinates $x^\mu = (t,r,x^A)$, with $x^A=(\theta,\phi)$ being
standard spherical coordinates, the Schwarzschild metric is 
\begin{eqnarray}
    g_{\mu\nu} &=&-\Big(1-\frac{2M}{r}\Big)t_{,\mu}t_{,\nu}
    +\Big(1-\frac{2M}{r}\Big)^{-1}r_{,\mu}r_{,\nu} +r^2q_{\mu\nu}(x^A) .
    \label{eq:Schwarz_std}
\end{eqnarray}
Here we use standard comma notation to denote partial derivatives,
e.g.  $f_{,\mu} =\partial_\mu f $,
and $q_{\mu\nu}(x^A)$ is the round unit sphere metric defined with
respect to the Cartesian coordinates
$x^i =r r^i$, $r^i(x^A) = (\sin\theta\cos\phi,\sin\theta\sin\phi, \cos\theta)$ so that
$$q_{\mu\nu}dx^\mu dx^\nu = \delta_{ij}r^i_{,A}r^j_{,B}dx^Adx^B
   = d\theta^2 +\sin^2\theta d\phi^2.
$$
Introduction of the ``tortoise'' coordinate $r^* = r +2M \ln(\frac{r}{2M}-1)$, with
$ r^*_{,\mu} = (rr_{,\mu})(r-2M)^{-1}$, gives
\begin{eqnarray}
    g_{\mu\nu} 
  &=&-\Big(1-\frac{2M}{r}\Big)(t_{,\mu}t_{,\nu} -r^*_{,\mu}r^*_{,\nu}) 
    +r^2q_{\mu\nu}(x^A) .
    \label{eq:ks0}
\end{eqnarray}
In terms of the retarded time $u=t-r^*$,
\numparts
\begin{equation}
    g_{\mu\nu} =-\Big(1-\frac{2M}{r}\Big)u_{,\mu}u_{,\nu}
    -u_{,\mu}r_{,\nu}- u_{,\nu}r_{,\mu}
    +r^2q_{\mu\nu}
    \label{eq:ks-}
\end{equation}
and in terms of the advanced time $v=t+r^*$, 
\begin{equation}
    g_{\mu\nu} =-\Big(1-\frac{2M}{r}\Big)v_{,\mu}v_{,\nu}
    +v_{,\mu}r_{,\nu} +v_{,\nu}r_{,\mu}
    +r^2q_{\mu\nu}.
    \label{eq:ks+}
\end{equation}
\endnumparts

The retarded time version of the Schwarzschild metric (\ref{eq:ks-})
has the Kerr-Schild form with Minkowski metric
$\eta^{(-)}_{\mu\nu}$,
\begin{equation}
    g_{\mu\nu} = \eta^{(-)}_{\mu\nu} 
    +\frac{2M}{r}k_\mu k_\nu , \quad k_\mu = -u_{,\mu}.
    \label{eq:mks-}
\end{equation}
Here, in the associated inertial coordinates
$x^{(-)\mu}=(t^{(-)},x^{(-)i})=(t^{(-)},x^{(-)},y^{(-)},z^{(-)})$,
with $t^{(-)}=u+r$ and $x^{(-)i} = r r^i(x^A)$,
\begin{equation}
  \eta^{(-)}_{\mu\nu} dx^{(-)\mu} dx^{(-)\nu} 
  =-dt^{(-)2} +\delta_{ij}dx^{(-)i} dx^{(-)j}  .
 \label{eq:outmink}
\end{equation}

Similarly, the advanced time version (\ref{eq:ks+})
has the Kerr-Schild form with
the background Minkowski metric
$\eta^{(+)}_{\mu\nu}$,
\begin{equation}
    g_{\mu\nu} = \eta^{(+)}_{\mu\nu} 
    +\frac{2M}{r} n_\mu n_\nu , \quad n_\mu = -v_{,\mu}
    \label{eq:mks+}
\end{equation}
where in the associated inertial coordinates
$x^{(+)\mu}
=(t^{(+)},x^{(+)i})=(t^{(+)},x^{(+)},y^{(+)},z^{(+)})$,
with $t^{(+)}=v-r^{(-)}$ and $x^{(+)i} =r r^i(x^A)$,
\begin{equation}
    \eta^{(+)}_{\mu\nu} dx^{(+)\mu} dx^{(+)\nu} 
 =-dt^{(+)2}+\delta_{ij}dx^{(+)i} dx^{(+)j}  .
 \label{eq:inmink}
\end{equation}
Note that the inertial time coordinates are related by
\begin{equation}
\label{eq:inertial_time_relation}
 t^{(+)}=t^{(-)}+ 4M \ln\Big(\frac{r}{2M}-1\Big) 
 \end{equation}
whereas the inertial spatial coordinates are related by
$x^{(+)i}=x^{(-)i}$. As a result, it is unambiguous to write
$x^{(+)i}=x^{(-)i}=x^i$ and
$r^{(+)}=r^{(-)} =r$, where $r^{(+)2}:=\delta_{ij}x^{(+)i}x^{(+)j}$
and $r^{(-)2}:=\delta_{ij}x^{(-)i}x^{(-)j}$.
However, the corresponding directional derivatives are related by 
\begin{equation}
\partial_{t^{(+)}} = \partial_{t^{(-)}}\, , \quad
    \partial_{x^{(+)i}} = \partial_{x^{(-)i}} -\frac{4M}{r-2M} \frac{x^i}{r}\partial_{t^{(-)}} 
\end{equation}
and
\begin{equation}
    \partial_{r^{(+)}} = \partial_{r^{(-)}} -\frac{4M}{r-2M} \partial_{t^{(-)}} .
\end{equation}
As will be seen, these transformations have important bearing
on the relation between the generators of the
BMS group at past and future null infinity.

In~\cite{boost}, we showed that the linearized memory
effect could be based upon the boosted
version of the advanced time Kerr-Schild metric (\ref{eq:mks+}).
In that linearized treatment, it was assumed that the boost was a Lorentz
symmetry of the Minkowski background. However,
this cannot be extended unambiguously to the nonlinear
case, where there are two distinct Minkowski backgrounds
$ \eta^{(-)}_{\alpha\beta}$ and $ \eta^{(+)}_{\alpha\beta}$
defined, respectively, by the retarded and advanced time Kerr-Schild
metrics (\ref{eq:mks-}) and (\ref{eq:mks+}).
Since the
metrics  (\ref{eq:mks-}) and (\ref{eq:mks+})  are algebraically
identical, it cannot be the choice of retarded or advanced
metric but the corresponding choice of boost that gives the essential result.

In spherical null coordinates, the Minkowski background 
metric (\ref{eq:outmink})
has the standard retarded time Bondi-Sachs form
at ${\mathcal I}^+$,
\begin{equation}\label{eq:BS_mink_ret}
  \eta^{(-)}_{\mu\nu}dx^\mu dx^\nu
  =-du^2 -2du dr +r^2 q_{AB}dx^Adx^B
\end{equation}
and (\ref{eq:inmink}) has the standard advanced time
Bondi-Sachs form at ${\mathcal I}^-$,
\begin{equation}\label{eq:BS_mink_adv}
  \eta^{(+)}_{\mu\nu}dx^\mu dx^\nu
  =-dv^2 +2dv dr +r^2q_{AB}dx^Adx^B.
\end{equation}
(See~\cite{bs_scolar} for a review of the Bondi-Sachs formalism.) 
These Minkowski line elements transform into each other under
the Minkowski space relation $u=v-2r$ but not
under the Schwarzschild relation between retarded
and advanced time $u=v-2r^*$.
The retarded and advanced Minkowski metrics
\eref{eq:outmink} and \eref{eq:inmink}  are
related by
\begin{eqnarray}
 \eta^{(+)}_{\mu\nu} &=&  \eta^{(-)}_{\mu\nu} 
    +\frac{2M}{r}\Big( k_{\mu}  k_{\nu}
    -  n_{\mu}  n_{\nu}\Big)
        \nonumber\\
   &=&\eta^{(-)}_{\mu\nu} 
    -\frac{4M }{r-2M} (u_{,\mu}r_{,\nu}+u_{,\nu}r_{,\mu})
    -\frac{8Mr}{(r-2M)^2}r_{,\mu}r_{,\nu} .
      \label{eq:eta}
\end{eqnarray}
Because of the non-vanishing $ r_{,\mu} r_{,\nu}$
term in (\ref{eq:eta}), although $\eta^{(+)}_{\mu\nu} $
has the advanced time Bondi-Sachs form
\eref{eq:BS_mink_adv} near
${\mathcal I}^-$
it does not have the retarded time Bondi-Sachs form
near ${\mathcal I}^+$. The reverse is true of  $\eta^{(-)}_{\alpha\beta} $,
which has the retarded time
Bondi-Sachs form  \eref{eq:BS_mink_ret} near ${\mathcal I}^+$
but not the advanced time form near ${\mathcal I}^-$.
This leads to a non-trivial difference between
the boosts $B^{(-)}$ and $B^{(+)}$, with
generators $B_{x^{(-)i}}$ and $B_{x^{(+)i}}$,
which are symmetries of $\eta^{(-)}_{\alpha\beta}$
and $\eta^{(+)}_{\alpha\beta}$, respectively.

To be specific, consider a boost in the $z^{(-)}$-direction intrinsic
to $\eta^{(-)}_{\mu\nu}$ with generator $B_{z^{(-)}}
= z^{(-)} \partial_{t^{(-)}} + t^{(-)}\partial_{z^{(-)}  }$.
In retarded spherical null coordinates
\begin{equation}
    \partial_{t^{(-)}} =\partial_u \,, \quad
    \partial_{z^{(-)}}  = -\cos\theta(\partial_u -\partial_{r^{(-)}})
     -\frac{\sin\theta}{r}\partial_\theta.
\end{equation}
This leads to the retarded time dependence
\begin{equation}
     B_{z^{(-)}} =-u\cos\theta \partial_u +(u+r)\cos\theta\partial_{r^{(-)}}
          -(\frac{u}{r} +1)\sin\theta\partial_\theta,
      \label{eq:bm}
\end{equation}
which has the proper asymptotic behavior to be the generator of a
BMS boost symmetry
at ${\mathcal I}^+$. 

However, expressed in terms of advanced null coordinates,
using $u=v-2r^*$, 
\begin{eqnarray}
      B_{z^{(-)}} &=&-(v-2r^*)\cos\theta \partial_v 
      +(2\partial_r r^*)(v-2r^*+r) \cos\theta \partial_v
      \nonumber \\
      &+&(v-2r^*+r)\cos\theta\partial_{r^{(+)}} -\left(\frac{v-2r^*}{r} +1\right)\sin\theta\partial_\theta
      \nonumber \\
   &=&\left\{\frac{(r+2M)[v-4M \ln(\frac{r}{2M} -1)]
    -4Mr}{r-2M} \right \} \cos\theta \partial_v  \nonumber \\
   &+&\left[v-r-4M \ln\left(\frac{r}{2M} -1\right)\right](\cos\theta\partial_{r^{(+)}}
   -\frac{1}{r}\sin\theta\partial_\theta).  
   \label{eq:bzm} 
\end{eqnarray}
Here, because the $\partial_v$ coefficient goes to infinity as
$ \ln(\frac{r}{2M} -1)$ for large $r$, $B_{z^{(-)}}$ 
generates a singular transformation at ${\mathcal I}^-$. 

Now consider the corresponding boost in the $z^{(+)}$-direction intrinsic
to $\eta^{(+)}_{\mu\nu}$ with generator
$B_{z^{(+)}}= z^{(+)} \partial_{t^{(+)}} + t^{(+)}\partial_{z^{(+)}}$. In advanced null coordinates,
\begin{equation}
     \partial_{t^{(+)}} =\partial_v ,  \quad
     \partial_{z^{(+)}} = \cos\theta(\partial_v +\partial_{r^{(+)}})
     -\frac{\sin\theta}{r}\partial_\theta.
\end{equation}
This leads to the advanced time dependence
\begin{equation}
      B_{z^{(+)}}=v\cos\theta \partial_v +(v-r)\cos\theta\partial_{r^{(+)}}
       -\left(\frac{v}{r} -1\right)\sin\theta\partial_\theta ,
      \label{eq:bp}
\end{equation}
which has the proper asymptotic behavior to be the generator of
a BMS boost symmetry at ${\mathcal I}^-$.
However, expressed in terms of retarded null coordinates
\begin{eqnarray}
      B_{z^{(+)}} 
   &=&-\left\{\frac{(r+2M)[u+4M \ln(\frac{r}{2M} -1)]
    +4Mr}{r-2M} \right \} \cos\theta \partial_u  \nonumber \\
   &+&\left[u+r+4M \ln\left(\frac{r}{2M} -1\right)\right](\cos\theta\partial_{r^{(-)}}
   -\frac{1}{r}\sin\theta\partial_\theta),
   \label{eq:bzp} 
\end{eqnarray}
which generates a singular transformation at  ${\mathcal I}^+$.
As will be seen in the next section, these unexpected gauge singularities
of $B_{z^{(-)}}$ at  ${\mathcal I}^-$  and
$B_{z^{(+)}}$ at  ${\mathcal I}^+$ do not affect calculation
of the radiation memory. 

%%%%%%%%%%%%%%%%%%%%%%%%%%%

\section{Boosts  and radiation memory}
\label{sec:mem}

In retarded null coordinates $x^\mu=(u,r,x^A)$, where the
angular coordinates $x^A=(\theta,\phi)$, are constant along the
outgoing null rays and $r$ is an areal coordinate which varies along the null rays,
the metric takes the Bondi-Sachs form
\begin{equation}
\label{BS_metric}
\fl g_{\mu\nu}dx^\mu dx^\nu = -\frac{V}{r}e^{2\beta} du^2-2 e^{2\beta}dudr 
+r^2h_{AB}\Big(dx^A-U^Adu\Big)\Big(dx^B-U^Bdu\Big) .
\end{equation}
The choice of areal coordinate $r$ and the
choice  $x^A=(\theta,\phi)$ as angular coordinates requires
\begin{equation}
      \det [h_{AB}] =  \det [q_{AB}] =\sin^2 \theta .
      \label{eq:det}
\end{equation}
As a result, the conformal 2-metric $h_{AB}$ has only two degrees of freedom,
which encode the two polarization modes of a
gravitational wave.

In the neighborhood of ${\mathcal I}^+$, asymptotic flatness allows
the construction of inertial coordinates such that
the metric approaches the Minkowski metric,
\begin{equation}
           \frac{V}{r}  =1 +O(1/r)\, , \quad 
    \beta = O(1/r^2) \, , \quad
    U^A =-\frac{1}{2r^2}\eth_E c^{EA}+O(1/r^3)
\end{equation}
and
\begin{equation}    
      h_{AB} = q_{AB} + c_{AB}/r +O(1/r^2).
\end{equation}
Here $\eth_A$ is the covariant derivative with respect to
$q_{AB}$ and the determinant condition (\ref{eq:det})
implies that $c_{AB}(u,x^C)$ is traceless, $q^{AB}c_{AB}=0$.
Evaluation of the geodesic deviation equation in the linearised limit
of the Bondi-Sachs metric shows that $\sigma_{AB} = \frac{1}{2}c_{AB}$ 
is the $O(1/r)$ strain tensor 
of the gravitational radiation.

The gravitational wave memory effect is determined by
the change in the radiation strain between
infinite future and past retarded time,
\begin{equation}\label{def_memory}
       \Delta  \sigma_{AB}(x^C): 
       = \sigma_{AB}(u=\infty,\theta,\phi) -  \sigma_{AB}(u=-\infty,\theta,\phi).
\end{equation}    
This produces a net displacement in the relative angular position of
distant test particles,\footnote{Note \eref{eq:displacment} corrects a missing
$1/r$ factor in the corresponding equation in~\cite{sky_pattern}.} 
\begin{equation}
\label{eq:displacment}
      \Delta (  x_2^A -  x_1^A) = \frac{1}{r}
      (  x_2^C -  x_1^C)q^{AB}  \Delta \sigma_{BC} .
\end{equation} 

A compact way to describe the radiation is in terms of
a complex polarization dyad $q_A$ satisfying 
\begin{equation}
          q _{AB} = \frac{1}{2}(q_A \bar q_B +\bar q_A  q_B), \quad
           q^A \bar q_A =2, 
           \quad q^A q_A =0.
\end{equation}
For the standard form of the unit sphere metric in
spherical coordinates $x^A=(\theta, \phi)$, we set
$q^A\partial_A =  \partial_\theta +(i/\sin\theta)\partial_\phi $.
In the associated  inertial Cartesian coordinates,
the dyad $q^A$ has components
$q^\mu = r Q^\mu = (0,rQ^i)$, where
$$Q^i = r^i_{,A}q^A =
( \cos\theta\cos\phi-i\sin\phi,  \cos\theta\sin\phi+i\cos\phi,-\sin\theta)$$
and $\delta_{ij}Q^i \bar Q^j =2$.
The dyad decompsition
\begin{equation}
\label{ }
\sigma_{AB}=\frac{1}{4}
  \Big[(q^Eq^F\sigma_{EF})\bar q_A\bar q_B 
  + (\bar q^E\bar q^F\sigma_{EF})q_Aq_B\Big]\;\;,
\end{equation}
leads to the spin-weight-2 representation of the strain,
\begin{equation}\label{def_rad_mem_spin0}
     \sigma:=\frac{1}{2}q^A q^B \sigma_{AB}.
\end{equation}
Note that $\sigma$ also corresponds to the leading $(r^{-2})$ coefficient
of the shear of the null hypersurfaces
$u=const$. Its retarded time derivative
$N(u,x^A):=\partial_u\sigma(u, x^A)$ is the Bondi news function.

The shear-free property of the Schwarzschild metric in its
rest frame implies that $\sigma=0$. For a transition from an initially
static Schwarzschild frame to a final boosted state, the
resulting spin-weighted radiation memory is then
\begin{equation}
     \Delta \sigma(x^C) = \sigma(u=\infty, x^C)- \sigma(u=-\infty,x^C) ,
     \label{eq:memeff}
\end{equation}
where $\sigma(u=\infty, x^C)$ is the radiation
strain of the final boosted state and initially $\sigma(u=-\infty, x^C) =0$.

Under the retarded time transformation $u\rightarrow u
+\alpha(x^A)$,
which corresponds to the supertranslation freedom in the
BMS group~\cite{Sachs_BMS}, the asymptotic strain has the gauge freedom
\begin{equation}
\sigma(u,x^A) \rightarrow \sigma(u,x^A)+\eth^2 \alpha(x^A),
\label{eq:super}
\end{equation}
where $\eth$ is the Newman-Penrose spin-weight
raising operator~\cite{BMS2}.
Since the finiteness of the radiative
mass loss requires that the news function $N=\partial_u \sigma$
vanish as $u\rightarrow \pm \infty$, the strain $\sigma$
can be gauged to zero either as $u\rightarrow \infty$ or
$u\rightarrow -\infty$.
The memory effect $\Delta \sigma$ \eref{def_memory} is gauge
invariant but determines a supertranslation
$\alpha(x^A)$ according to
$$\eth^2 \alpha(x^A) =\Delta \sigma(x^A),
$$
which relates the strains
at $u=\pm \infty$. The energy flux of the
radiation is given by the absolute square, $N \bar N  $,
of the Bondi news function
$N$, which
is also gauge invariant. 
If the memory effect
(\ref{eq:memeff}) is non-zero then there must be
intervening radiation.

These attributes of ${\mathcal I}^+$ have corresponding
attributes at ${\mathcal I}^-$. In particular, the
outgoing radiation strain $\sigma(u,x^A)$ has as its
analogue an ingoing radiation strain $\Sigma(v,x^A)$.
In analogy with (\ref{eq:memeff}), the gravitational wave memory at
${\mathcal I}^-$ due to ingoing radiation is 
\begin{equation}
     \Delta \Sigma(x^C) = \Sigma(v=\infty, x^C)
     - \Sigma(v=-\infty,x^C) .
     \label{eq:pastmemeff}
\end{equation}
If there is no ingoing radiation, as required in the
linearized case by a retarded solution, then 
$\Delta \Sigma(x^C) =0$.

Of the BMS transformations, only the supertranslations
(\ref{eq:super}) affect the radiation strain.
As  shown in Sec.~\ref{sec:advret},
a $B^{(-)}$ boost is a BMS boost symmetry at
${\mathcal I}^+$ so that it does not introduce
outgoing radiation memory $\Delta \sigma$.
Conversely, a $B^{(+)}$ boost is a BMS boost symmetry
at ${\mathcal I}^-$ so that it does not introduce
ingoing radiation memory $\Delta \Sigma$.
These results are explicitly demonstrated below.

\subsection{Effect of a $B^{(-)}$ boost}
\label{sec:bminus}

Consider first the transition from a static Kerr-Schild-Schwarzschild metric to the
$B^{(-)}$ boosted version
with 4-velocity $v^\mu = \Gamma(1, V^i)$,
where $\Gamma = (1 -\delta_{ij}V^iV^j)^{-1/2}$. For a $B^{(-)}$ boost,
$\eta^{(-)}_{\mu\nu} \rightarrow \eta^{(-)}_{\mu \nu}$.
The boosted version of the static retarded time
Kerr-Schild-Schwarzschild metric
(\ref{eq:mks-}),  can be obtained by
the further substitutions 
\begin{equation}
\fl \partial_\mu t^{(-)} \rightarrow -v_\mu, \quad
    r^2  \rightarrow R^{(-)2} = x^{(-)}_\mu x^{(-)\mu} +(x^{(-)}_\mu v^\mu)^2, \quad
k_\mu \rightarrow K_\mu = v_\mu+R^{(-)} _\mu,
\end{equation}
where
\begin{equation}
       \partial_\mu r \rightarrow R^{(-)} _\mu =
       \frac{1}{ R^{(-)}}(x^{(-)}_\mu
       + v_\mu x^{(-)}_\nu v^\nu).
 \end{equation}
The boosted metric is
\begin{equation}
    g^{(B^-)}_{\mu\nu} = \eta^{(-)}_{\mu\nu} 
    +\frac{2M}{R^{(-)}}K_\mu K_\nu .
\end{equation}
This Lorentz covariant transformation reduces to 
the rest frame expression when $V^i=0$.

In order to calculate the resulting radiation strain,
we note that $q^\mu q^\nu \eta^{(-)}_{\mu\nu}=0$ 
and $q^\mu x_\mu$ =0 so that
$$\sigma^{(B^-)}=\frac{1}{4} rq^\mu q^\nu g^{(B^-)}_{\mu\nu} \big|_{\mathcal{I}^+} 
=\frac{M r}{2R^{-}}(q^\mu K_\mu)^2\big|_{\mathcal{I}^+},
$$
where $q^\mu K_\mu = q^\mu v_\mu (1 + R^{-1} x^{(-)}_\nu v^\nu)$.

For the limit at
${\mathcal I}^+$, in retarded null coordinates
\numparts
\begin{eqnarray}
R^{(-)2}  &=& r^2\Big[-\frac{u^2}{r^2}-\frac{2u}{r}
  + \Gamma^2\Big(1-\frac{x _iV^i}{r} + \frac{u}{r}\Big)^2\Big]\;,\\
 x^{(-)}_\nu v^\nu &=& \Gamma(-u-r + x_i V^i) \;,
\end{eqnarray}
\endnumparts
so that 
\numparts
\begin{eqnarray}
 \lim_{\stackrel{r\rightarrow\infty}{u=const}}\frac{R^{(-)}}{r}
    &=&\Gamma(1-\frac{V^i x_i}{r}) \;, \\
 \lim_{\stackrel{r\rightarrow\infty}{u=const}}
 \frac{ x^{(-)}_\nu v^\nu}{r}&=& -\Gamma(1-\frac{V^i x_i}{r}) \;.
\end{eqnarray}
\endnumparts
Consequently, 
\begin{equation}
 \lim_{\stackrel{r\rightarrow\infty}{u=const}}
 \frac{x^{(-)}_\nu v^\nu}{R^{(-)}} = -1 
\end{equation}
and 
\begin{equation}
 \lim_{\stackrel{r\rightarrow\infty}{u=const}}
   q^\mu K_\mu =0.
\end{equation}
Therefore $\sigma^{(B^-)}(u,x^C)=0$ and in particular
$\sigma^{(B^-)}(u=\infty, x^C)=0$.  So, as expected
from the BMS property of the $B^{(-)}$ boost at ${\mathcal I}^+$, it produces no radiation memory at ${\mathcal I}^+$.
Now consider the boosted strain on ${\mathcal I}^-$,
\begin{equation}
    \Sigma^{(B^-)} 
      = \frac{r}{4}q^\mu q^\nu g^{(B^-)}_{\mu\nu}    |_{{\mathcal I}^-}
   =\frac{Mr}{2R^{(-)}}(q^\mu v_\mu )^2
     \Big(1+\frac{x^{(-)}_\nu v^\nu}{R^{(-)}}\Big)^2
      \Big|_{{\mathcal I}^-} .
\end{equation} 
In order to calculate the limit at ${\mathcal I}^-$, for which 
$r\rightarrow \infty$ holding $v =t^{(+)} +r$ constant, we must
express $\Sigma^{(B^-)}$ as a function of  the unboosted advanced
coordinates $(v,r,x^A)$. Using \eref{eq:inertial_time_relation}, a straightforward calculation gives 
\begin{equation}
    x^{(-)}_\mu v^\mu = r \Gamma\Big[1-\frac{v}{r}
             +\frac{4M}{r}\ln(\frac{r}{2M}-1)
                       +r_i V^i\Big], 
\end{equation}
\begin{equation}
    x^{(-)}_\mu x^{(-)\mu}= r[v-4M \ln(\frac{r}{2M}-1)]
      [2-\frac{v}{r}+\frac{4M}{r} \ln(\frac{r}{2M}-1)] , 
\end{equation}
which leads to the limits 
\begin{equation}
 \lim_{\stackrel{r\rightarrow\infty}{v=const}}\frac{R^{(-)}}{r}
     =\Gamma(1+V^ir_i) , 
\end{equation}
\begin{equation}
 \lim_{\stackrel{r\rightarrow\infty}{v=const}}
     \frac{x^{(-)}_\mu v^\mu }{R^{(-)}}
    = \lim_{\stackrel{r\rightarrow\infty}{v=const}}
   \big[ \frac{r}{R^{(-)}} \big ]\big [\frac{x^{(-)}_\mu v^\mu }{r} \big]
    =1 .  
\end{equation}
We then obtain
\begin{equation}\label{eq:pastboostmem}
    \Sigma^{(B^-)} = \frac{2Mr}{R^{(-)}}(q^\mu v_\mu )^2
      |_{{\mathcal I}^-}
      =\frac{2M\Gamma (q^iV_i)^2}{1+V^ir_i} \;\;.
\end{equation} 
Consequently, for a non zero boost, the resulting radiation memory on ${\mathcal I}^-$
does not vanish, which requires  the existence of ingoing
radiation.

Thus the $B^{(-)}$ boost is inconsistent
with vanishing ingoing radiation and produces zero
radiation memory on  ${\mathcal I}^+$.
Both of these results contradict the linearized result based
upon the retarded Green function so that $B^{(-)}$ is not
the appropriate boost to model the memory effect. 

%%%%%%%%%%%%%%%%%%%%%%%

\subsection{Effect of a $B^{(+)}$ boost}

Consider now the
transition from a static to a $B^{(+)}$ boosted
version of the Kerr-Schild-Schwarzschild metric with 4-velocity
$v^\mu = \Gamma(1, V^i)$. For the $B^{(+)}$ boost,
$\eta^{(+)}_{ab} \rightarrow \eta^{(+)}_{ab}$.
With respect to the advanced time version of the static
Kerr-Schild-Schwarzschild metric (\ref{eq:mks+}),
the boosted version can be obtained by
the further substitutions 
\begin{equation}
\fl   \partial_\mu t^{(+)} \rightarrow -v_\mu, \quad
  r^2  \rightarrow R^{(+)2} =   x^{(+)}_\mu x^{(+)\mu}
     +(x^{(+)}_\mu v^\mu)^2 , \quad
    n_\mu  \rightarrow N_\mu
     = v_\mu-R^{(+)} _\mu,
\end{equation}
where
\begin{equation}
       \partial_\mu r \rightarrow R^{(+)} _\mu =
       \frac{1}{ R^{(+)}}(x^{(+)}_\mu
       + v_\mu x^{(+)}_\nu v^\nu).
 \end{equation}
The boosted metric is
\begin{equation}
    g^{(B^+)}_{\mu\nu} = \eta^{(+)}_{\mu\nu} 
    +\frac{2M}{R^{(+)}}N_\mu N_\nu 
\end{equation}    
with the corresponding boosted strain on ${\mathcal I}^-$ given
by
\begin{equation}
   \fl\quad
    \Sigma^{(B^+)}  
    = \frac{r}{4}q^\mu q^\nu g^{(B^+)}_{\mu\nu}\Big|_{{\mathcal I}^-}
    = \frac{M r}{ 2R^{(+)}}(q^\mu N_\mu )^2 \Big|_{{\mathcal I}^-}
               =\frac{Mr}{2R^{(+)}}(q^\mu v_\mu )^2
           \Big(1-\frac{x^{(+)}_\nu v^\nu}{R^{(+)}}\Big)^2
           \Big|_{{\mathcal I}^-}. \quad
\end{equation} 
The calculation of the limit proceeds in a time reversed
sense as in Sec.~\ref{sec:bminus}.
 
In advanced null coordinates
\numparts
\begin{eqnarray}
R^{(+)2}  &=& r^2\Big[-\frac{v^2}{r^2}+\frac{2v}{r}
  + \Gamma^2\Big(1+\frac{x _iV^i}{r} - \frac{v}{r}\Big)^2\Big]\;,\quad \\
 x^{(+)}_\nu v^\nu &=& \Gamma(-v+r + x_i V^i) \;,
\end{eqnarray}
\endnumparts
so that 
\numparts
\begin{eqnarray}
 \lim_{\stackrel{r\rightarrow\infty}{v=const}}\frac{R^{(+)}}{r}
    &=&\Gamma(1+\frac{V^i x_i}{r}) \;,\qquad  \\
 \lim_{\stackrel{r\rightarrow\infty}{v=const}}
 \frac{ x^{(+)}_\nu v^\nu}{r}&=& \Gamma(1+\frac{V^i x_i}{r}) \;.
\end{eqnarray}
\endnumparts
Consequently, 
\begin{equation}
 \lim_{\stackrel{r\rightarrow\infty}{v=const}}
 \frac{x^{(+)}_\nu v^\nu}{R^{(+)}} 
  = \lim_{\stackrel{r\rightarrow\infty}{v=const}}
 \big[\frac{r}{R^{(+)}} \bi]\big[ \frac{x^{(+)}_\nu v^\nu}{r}\big]
 = 1 \qquad
\end{equation}
and therefore $\Sigma^{(B^+)}(v,x^C)=0$. So, as expected from the BMS property of the $B^{(+)}$ boost
at ${\mathcal I}^-$,
there is no radiation memory at ${\mathcal I}^-$.
It is thus consistent to set the free
characteristic initial data $\Sigma$ to zero on
${\mathcal I}^-$ so that there is no ingoing radiation.

Now consider the boosted strain on ${\mathcal I}^+$,
\begin{equation}
 \sigma^{(B^+)}
   = \frac{r}{4}q^\mu q^\nu g^{(B^+)}_{\mu\nu}
          |_{{\mathcal I}^+}
   =\frac{Mr}{2R^{(+)}}(q^\mu v_\mu )^2
     \Big(1-\frac{x^{(+)}_\nu v^\nu}{R^{(+)}}\Big)^2
      \Big|_{{\mathcal I}^+} .
\end{equation} 
In order to calculate the limit at ${\mathcal I}^+$, for which 
$r\rightarrow \infty$ holding $u =t^{(-)} -r$ constant, we must
express $\sigma^{(B^+)}$ as a function of  the unboosted retarded
coordinates $(u,r,x^A)$. A straightforward calculation gives
\begin{equation}
    x^{(+)}_\mu v^\mu = -r \Gamma\Big[1+\frac{u}{r}
             +\frac{4M}{r}\ln(\frac{r}{2M}-1)
                       -r_i V^i\Big],
\end{equation}
\begin{equation}
    x^{(+)}_\mu x^{(+)\mu}= -r[u+4M \ln(\frac{r}{2M}-1)]
      [2+\frac{u}{r}+\frac{4M}{r} \ln(\frac{r}{2M}-1)] , 
\end{equation}
which leads to the limits 
\begin{equation}
 \lim_{\stackrel{r\rightarrow\infty}{u=const}}\frac{R^{(+)}}{r}
     =\Gamma(1-V^ir_i) ,
\end{equation}
\begin{equation}
 \lim_{\stackrel{r\rightarrow\infty}{u=const}}
     \frac{x_\nu v^\nu}{R^{(+)}}
  =   \lim_{\stackrel{r\rightarrow\infty}{u=const}}
   \frac{-\Gamma r\Big[1-V^i r_i +\frac{u}{r}
         +\frac{4M}{r}\ln(\frac{r}{2M}-1)\big]}{R^{(+)}}   =-1 .
\end{equation}
We then obtain 
\begin{equation}
    \sigma^{(B^+)} =\frac{2Mr}{R^{(+)}}(q^\mu v_\mu )^2
      |_{{\mathcal I}^+} = \frac{2M\Gamma}{(1-r_i V^i)}(q^i V_i )^2.
\end{equation} 
The resulting radiation memory due to the ejection of a
Schwarzschild body is
\begin{equation}
  \Delta \sigma^{(B^+)} = \frac{2M\Gamma}{1-r_i V^i}(q^i V_i )^2.
\end{equation}
This is in exact agreement with the linearized result.

%%%%%%%%%%%%%%%%%

\section{Discussion}
\label{sec:discuss}

We have shown that the boost symmetry $B^{(+)}$  of the
Minkowski background $\eta^{(+)}_{\mu\nu} $
of the ingoing Kerr-Schild version of the Schwarzschild metric
leads to a nonlinear model for determining the memory effect due to
the ejection of a massive particle. An initially stationary
Kerr-Schild-Schwarzschild metric followed by
an accelerating interval which produces radiation and
leads to a final $B^{(+)}$ boosted state
is consistent with the absence of ingoing radiation
and produces outgoing radiation in agreement with
the linearized memory effect obtained from a retarded
solution. The corresponding results for a
 $B^{(-)}$  boost of the
Minkowski background $\eta^{(-)}_{\mu\nu}$
produces results expected  in the linearized limit
from the use of an advanced Green function.

In~\cite{boost}, we have given an analysis of how radiation memory
affects angular momentum conservation. In a non-radiative regime, a preferred
Poincar{\'e} subgroup can be picked out from the BMS group.
This difference $\Delta \sigma$ between
initial and final radiation strains induces the supertranslation shift (\ref{eq:super})
between the preferred Poincar{\'e} groups at $u=\pm \infty$.
The rotation subgroups associated
with the initial and final Poincar{\'e} groups differ by a supertranslation. 
As a result, the corresponding components of angular momentum
intrinsic to the initial and final states differ by supermomenta.
This complicates the interpretation
of angular momentum flux conservation laws.
There might be a distinctly general relativistic mechanism
for angular momentum loss.
This is a ripe area for numerical investigation.

In prescribing
initial data for the numerical simulation of binary
black holes using superimposed Kerr-Schild metrics~\cite{ks1,ks2},
$B^{(+)}$ is used to induce the orbital motion.
Although $B^{(+)}$ has a logarithmic
singularity (\ref{eq:bzp})  at  ${\mathcal I}^+$, this is a pure gauge effect
which does not show up in the memory effect
measured by the change in asymptotic strain $\Delta \sigma$ but it could introduce spurious effects
in the prescription of binary black hole initial data.
Whether this adversely affects the asymptotic gauge behavior
of the data deserves further study.

The model presented here provides a scheme for
studying these issues. Although our example of a
transition from a asymptotically stationary to boosted state is highly idealized,
the chief criterion for the model is that,
to an asymptotic approximation, the far field
behavior of the  initial and final states consist of the
Kerr-Schild superposition of distant Schwarzschild bodies.
The model is also applicable to an initial state whose far field is
a superposition of boosted Schwarzschild bodies which,
after some dynamic, radiative process, coalesce to form
a boosted Kerr black hole. Of course, the intermediate radiative epoch,
which determines the final mass and velocity, 
must be treated by numerical methods. The
Kerr-Schild model offers a
framework for interpreting such results.

%%%%%%%%%%%%%%%%%%%%

\ack
We thank the AEI in Golm  for hospitality during this project.
T.M. appreciates  support  from  C. Malone and the University of Cambridge.  
J.W. was supported by NSF grant PHY-1505965 to the University of Pittsburgh.

%%%%%%%%%%%%%%%%%%%%

\end{document}